\documentclass[12pt,preprint]{aastex}

\usepackage{float}
\usepackage{graphicx}
\usepackage{epsfig}
\usepackage[perpage,symbol*]{footmisc}
\usepackage{natbib}

\def\nat{Nature}
\def\apj{ApJ}

\def\apjl{ApJ}
\def\apjs{ApJ}

\def\mnras{MNRAS}

\def\aap{A\&A}

\def\pasa{PASA}

\shorttitle{Single-pulse observations of PSR~J1745$-$2900}
\shortauthors{Z. Yan et al.}

\begin{document}


\title{Single-pulse radio observations of the Galactic Center magnetar PSR~J1745$-$2900}


\author{Zhen~Yan \thanks{yanzhen@shao.ac.cn} \altaffilmark{1,7},
  Zhi-Qiang~Shen\altaffilmark{1,7}, Xin-Ji~Wu\altaffilmark{2},
  R.~N.~Manchester\altaffilmark{3}, P.~Weltevrede\altaffilmark{4},
  Ya-Jun~Wu\altaffilmark{1,7}, Rong-Bing~Zhao\altaffilmark{1,7},
  Jian-Ping~Yuan\altaffilmark{5,7}, Ke-Jia~Lee\altaffilmark{6},
  Qing-Yuan~Fan\altaffilmark{1,7}, Xiao-Yu~Hong\altaffilmark{1,7},
  Dong-Rong~Jiang\altaffilmark{1,7}, Bin~Li\altaffilmark{1,7},
  Shi-Guang~Liang\altaffilmark{1,7}, Quan-Bao~Ling\altaffilmark{1,7},
  Qing-Hui~Liu\altaffilmark{1,7}, Zhi-Han~Qian\altaffilmark{1,7},
  Xiu-Zhong~Zhang\altaffilmark{1,7}, Wei-Ye~Zhong\altaffilmark{1,7},
  Shu-Hua~Ye\altaffilmark{1,7}}


\altaffiltext{1}{Shanghai Astronomical Observatory, Chinese Academy of
  Sciences, Shanghai 200030, China} \altaffiltext{2}{Department of
  Astronomy, Peking University, Beijing 100871, China}
\altaffiltext{3}{CSIRO Astronomy and Space Science, PO Box 76, Epping
  NSW 1710, Australia} \altaffiltext{4}{Jodrell Bank Centre for
  Astrophysics, School of Physics and Astronomy, University of
  Manchester, Manchester M13 9PL, UK } \altaffiltext{5}{Xinjiang
  Astronomical Observatory, Chinese Academy of Sciences, Urumqi
  830011, China} \altaffiltext{6} {Kavli Institute for Astronomy and
  Astrophysics, Peking University, Beijing 100871, China}
  \altaffiltext{7}{Key Laboratory of Radio Astronomy,
  Chinese Academy of Sciences, China}


\begin{abstract}
In this paper, we report radio observations of the Galactic Center
magnetar PSR~J1745$-$2900 at six epochs between June and October,
2014. These observations were carried out using the new Shanghai Tian
Ma Radio Telescope at a frequency of 8.6~GHz. Both the flux
density and integrated profile of PSR~J1745$-$2900 show dramatic
changes from epoch to epoch showing that the pulsar was in its
``erratic'' phase. On MJD~56836, the flux density of this magnetar was
about 8.7~mJy, which was ten times large than that reported at the
time of discovery, enabling a single-pulse
analysis. The emission is dominated by narrow ``spiky'' pulses which
follow a log-normal distribution in peak flux density. From 1913
pulses, we detected 53 pulses whose peak flux density is ten times
greater than that of the integrated profile. They are concentrated in
pulse phase at the peaks of the integrated profile. The pulse widths
at the 50\% level of these bright pulses was between 0.2$^\circ$ to
0.9$^\circ$, much narrower than that of integrated profile
($\sim$12$^\circ$). The observed pulse widths may be limited by
interstellar scattering. No clear correlation was found between the
widths and peak flux density of these pulses and no evidence was found
for subpulse drifting. Relatively strong spiky pulses are also detected
in the other five epochs of observation, showing the same
properties as that detected in MJD~56836. These strong spiky
pulses cannot be classified as ``giant'' pulses but are more closely related
to normal pulse emission.
\end{abstract}





\section{Introduction}\label{sec:intro}
Observations of a pulsar closely orbiting Sgr~A*, the supermassive
black hole at the Galactic Center (GC), could provide a variety of
important information about the pulsar itself, the black hole, the
interstellar medium around the GC and, if close enough, relativistic
orbital dynamics. But, experience has shown that radio searches for
pulsars near the GC are really challenging. Many searches have been
made \citep{kkl00, jkl06, dcl09, mkf10, ekk13} but no pulsars were
discovered within a few arcmin of Sgr~A*.  The closest separations
between pulsars discovered by these projects and Sgr A* are about
10-15 arcmin.

PSR~J1745$-$2900, which was serendipitously discovered, is the only
pulsar known within a few arcmin from Sgr~A*. It was first detected as
an X-ray flare thought to be from Sgr A* by Swift \citep{kbk13}. Follow-up
observations with the NuSTAR X-ray Observatory detected periodic intensity
variations with a period of 3.76~s \citep{mgz+13}. It is confirmed as a
magnetar by subsequent X-ray timing observations with NuStar and Swift \citep{kab+14}.
Unlike ordinary pulsars, most of which can only be detected in the radio band,
very few magnetars have been confirmed with radio pulsations. Of the 28 magnetars
and candidates previously known, only three of them have detectable radio
pulsations \citep{ols14}. The radio pulsations from PSR J1745$-$2900 were
detected by several large radio telescopes, which makes it the fourth magnetar
known with radio pulsations \citep[e.g.,][]{efk13}.

Astrometry measurements for PSR~J1745$-$2900 with the VLBA indicate that
its projected separation from Sgr~A* is as small as 0.097~pc \citep{bdd14},
giving it the potential to contribute to the important studies mentioned
above. PSR~J1745$-$2900 shows a flat
radio spectrum and a high degree of polarized emission, according to
observations at frequencies ranging from 1.4 to 20~GHz.  The derived
dispersion measure (DM) is $1778\pm3$~$\rm{cm^{-3}~pc}$ \citep{efk13},
which makes it the highest-DM pulsar known. The Faraday Rotation
Measure (RM) of this pulsar is $-66960\pm 50$~rad m$^{-2}$. This
implies that the magnetic field strength is about 2.6~mG at the
radius of 0.12~pc from Sgr~A*, which is dynamically important for
the accretion of the black hole \citep{efk13,shj13}.  The
scatter-broadening time scale at 1~GHz
and the scatter-broadening spectral index fitted with multi-frequency
observations are $1.3 \pm 0.2$~s and $-3.8 \pm 0.2$ respectively
\citep{sle+14}. This timescale is several orders of magnitude lower
than the prediction of the NE2001 model \citep{col02}. Judging from
this result, scattering effects are not the main reason for lack of
pulsar detections around Sgr~A*.  The intrinsic deficit in the
ordinary pulsar population in this area is proposed to be the
most likely reason for this \citep{deo14, chl14}.

Eight months of 8.7~GHz observations of PSR~J1745$-$2900 with the
Green Bank Telescope (GBT) between MJDs 56515 and 56845 \citep{lak15}
identified two main periods of activity. The first is characterized by
approximately 5.5 months (up to MJD~56726) of relatively stable
evolution in radio flux density, rotation and profile shape, while in
the second period these properties became highly variable.  The GBT
observations gave a single-pulse energy distribution that roughly
follows a log-normal distribution with an apparent high-energy
tail. Subpulse drifting was not found in their observations. No
sustained enhancement of the X-ray emission at the GC was detected by
the Swift on MJD~56910 \citep{drm14}. But, at present it is
not clear that this X-ray enhancement is related to PSR~J1745$-$2900
or Sgr~A*.

In this paper, we present the results of 8.6~GHz observations that
were carried out with the Shanghai Tian Ma Radio Telescope (TMRT). The
observation and data reduction procedures are presented in
\S\ref{sec:obs}. In \S\ref{sec:intprf} and \S\ref{sec:single} we
discuss the integrated profile and single-pulse properties,
respectively, and give further discussion about the results in
\S\ref{sec:concl}.

\section[]{Observations and data reduction procedures}\label{sec:obs}
\subsection{Introduction to the TMRT}
The TMRT is a new 65~m diameter fully-steerable radio telescope located in
the western suburbs of Shanghai, China. The first phase of construction
was finished in December 2013. Four cryogenically cooled receivers
covering the frequency ranges 1.25 -- 1.75~GHz, 2.2 -- 2.4~GHz, 4.0 -- 8.0~GHz
and 8.2 -- 9.0~GHz, respectively, are available. The highest frequency of the
TMRT will be 43~GHz. The telescope has an active surface control to compensate
for gravity deformation of the main reflector during tracking.

The digital backend system (DIBAS) of TMRT is an FPGA-based spectrometer
based upon the design of Versatile GBT Astronomical Spectrometer (VEGAS) with
pulsar modes that provide much the same capabilities as Green Bank Ultimate
Pulsar Processing Instrument (GUPPI) \citep{rdf09, bus12}. For pulsar
observations, DIBAS supports both pulsar
searching and on-line folding mode. Both coherent and incoherent
dedispersion observation modes are supported. For the incoherent
dedispersion observation mode, the bandwidth of each of digitizer
channel is up to 2~GHz. Since three pairs of digitizer (one
digitizer for each polarization) are currently available, a
maximum bandwidth of 6~GHz can be supported.  For the coherent
dedispersion observation mode the maximum bandwidth is 1~GHz,
limited by the computing power of the current high-performance
computer cluster. DIBAS supports full Stokes-parameter pulsar
observations which are written out in 8-bit PSRFITS format
\citep{hvm04}\footnote{See also
  http://www.atnf.csiro.au/research/pulsar/index.html?n=Main.Psrfits}.
In order to reduce the data rate in pulsar searching observations,
the observer can also choose to record Stokes-parameter I (total intensity)
only, which is obtained by summing the two polarization channels
after digitisation.

\subsection{Observations and data reduction}
\label{part:obs-data}
Observations of PSR~J1745$-$2900 were performed with the TMRT in 2014, June
to October using the incoherent pulsar searching observation mode. The
frequency range of our observations was 8.2 -- 9.0~GHz. The
full bandwidth was divided into 512 channels to allow off-line
dedispersion. Even though the DM of PSR~J1745$-$2900 is
1778~cm$^{-3}$~pc, at this high frequency, 512 channels are
sufficient, giving a time delay in each channel of only $\sim$~46.9
microsecond. The sampling interval was 131.07 microseconds. The exact
observing dates were 2014 June 28, 2014 July 24, 2014 Aug. 4, 2014
Sept. 11, 2014 Sept. 12 and 2014 Oct. 13, corresponding to MJDs of 56836,
56862, 56873, 56911, 56912 and 56943, respectively. The data recording
times for the six sessions were 120, 40, 25, 26, 60 and 90 min,
respectively. The first two epochs of observation presented
here occurred prior to the last epoch reported in \cite{lak15}. 

The Digital Signal Processing for Pulsars (DSPSR) program
\citep{stb11} was used to dedisperse and fold the data at the known
topocentric period using polynomial coefficients generated with TEMPO2
in its prediction mode \citep{hem06}. The profile data were written
out with 1024 phase bins per period in PSRFITS format. The pulse
broadening caused by interstellar scattering is about 3.7 millisecond
at 8.6~GHz \citep{bdd14} and so it is reasonable to use 1024 phase
bins across the 3.76~s pulse period. PSRCHIVE programs \citep{hvm04}
were used to do further data editing and processing.

At the time of observation, the diode that injects pulsed signals
into the front-end of the receiver had not been installed, so we
could not use the normal procedure to do the flux density calibration.
Previous studies have shown that the flux density of
pulsars is in general intrinsically stable over several years. Most
pulsars are in the weak scattering regime at 8.6~GHz \citep{stc90} and
so it is reasonable to use other normal pulsars as calibrators to
estimate the flux density of PSR~J1745$-$2900. The pulsars chosen as
the flux density calibrators were observed on the same day with
PSR~J1745$-$2900 using the same setups. There were no input power
adjustments between observations of calibrators and PSR~J1745$-$2900.

On MJD~56836, we observed five normal pulsars with known flux density
at 8.35~GHz with same setup as for
PSR~J1745$-$2900. Table~\ref{tab:fluxpsr} lists these pulsars and
gives their 8.35~GHz flux density measured at Effelsberg
\citep{msk13}, radio spectral index \citep{mgj94}, estimated flux
density at 8.6~GHz, the flux density in arbitrary units measured with
the TMRT, the average elevation angle of the TMRT during the
observation, the length of the observation and the scaling factor to
convert from TMRT units to mJy. Using the scaling factor of the
five normal pulsars, we obtain the weighted average scaling factor
and corresponding standard deviation. These are used to convert
the measured flux density of PSR~J1745$-$2900 into mJy units.
The effect of possible variations in the flux of the known pulsars
is negligible because of the stable intrinsic flux density of
calibrator pulsars, weak interstellar scattering at the band of
observation and the averaging effect of using multiple calibrators.
Therefore, we only use the standard deviation of scaling factor
to quantify the flux uncertainties of PSR~J1745$-$2900.
\begin{table*}[ht]
{\footnotesize
\caption{Flux densities of the calibration pulsars and of PSR J1745$-$2900 at MJD 56836}\label{tab:fluxpsr}
\begin{tabular}{lccccccc}
\hline
Name & S$_{8.35}$ & $\alpha$ & S$_{8.60}$ & S$_{\rm TM}$ & Elev. & T & Factor \\
 & (mJy)  & & mJy & (arb) & ($^\circ$) & (min) &  \\
\hline
B0329+54 & $1.99\pm0.40$  & $3.8\pm0.9$  & $1.87\pm0.40$ & 0.00079768  &  49.8 &   120 & $2344\pm501$ \\
B0809+74 & $0.64\pm0.13$  & $1.7\pm0.1$  & $0.61\pm0.13$ & 0.00024458  &  44.0 &   60  & $2494\pm532$ \\
B0823+26 & $0.86\pm0.17$  & $1.8\pm0.1$  & $0.82\pm0.16$ & 0.00033217  &  49.7 &   60  & $2469\pm482$ \\
B0950+08 & $1.09\pm0.22$  & $2.4\pm0.3$  & $1.02\pm0.21$ & 0.00051095  &  38.3 &   30  & $1996\pm410$ \\
B1133+16 & $0.78\pm0.16$  & $2.0\pm0.1$  & $0.74\pm0.15$ & 0.00034609  &  41.5 &   90  & $2138\pm433$ \\
\hline
{\it J1745$-$2900} & - & - & ${\it 8.75\pm0.81}$ & {\it 0.00388361} &{\it 29.3} & {\it 120} & ${\it 2254\pm208}$ \\
\hline
\end{tabular}}
\end{table*}

The estimated flux density of PSR~J1745$-$2900 at 8.6~GHz on MJD~56836 is
$8.75\pm0.81$~mJy. The GBT measurements carried out in the erratic
phase of PSR~J1745$-$2900 showed a mean flux density 11~mJy with a
standard deviation of 8.5~mJy and mean uncertainties of
$^{+3.1}_{-2.4}$~mJy \citep{lak15} and so the measured flux density of
PSR~J1745$-$2900 is reasonable.

\section{Integrated profile of PSR~J1745$-$2900}\label{sec:intprf}
\begin{figure*}[b]
\centering
\includegraphics[width=0.76\textwidth,angle=-90]{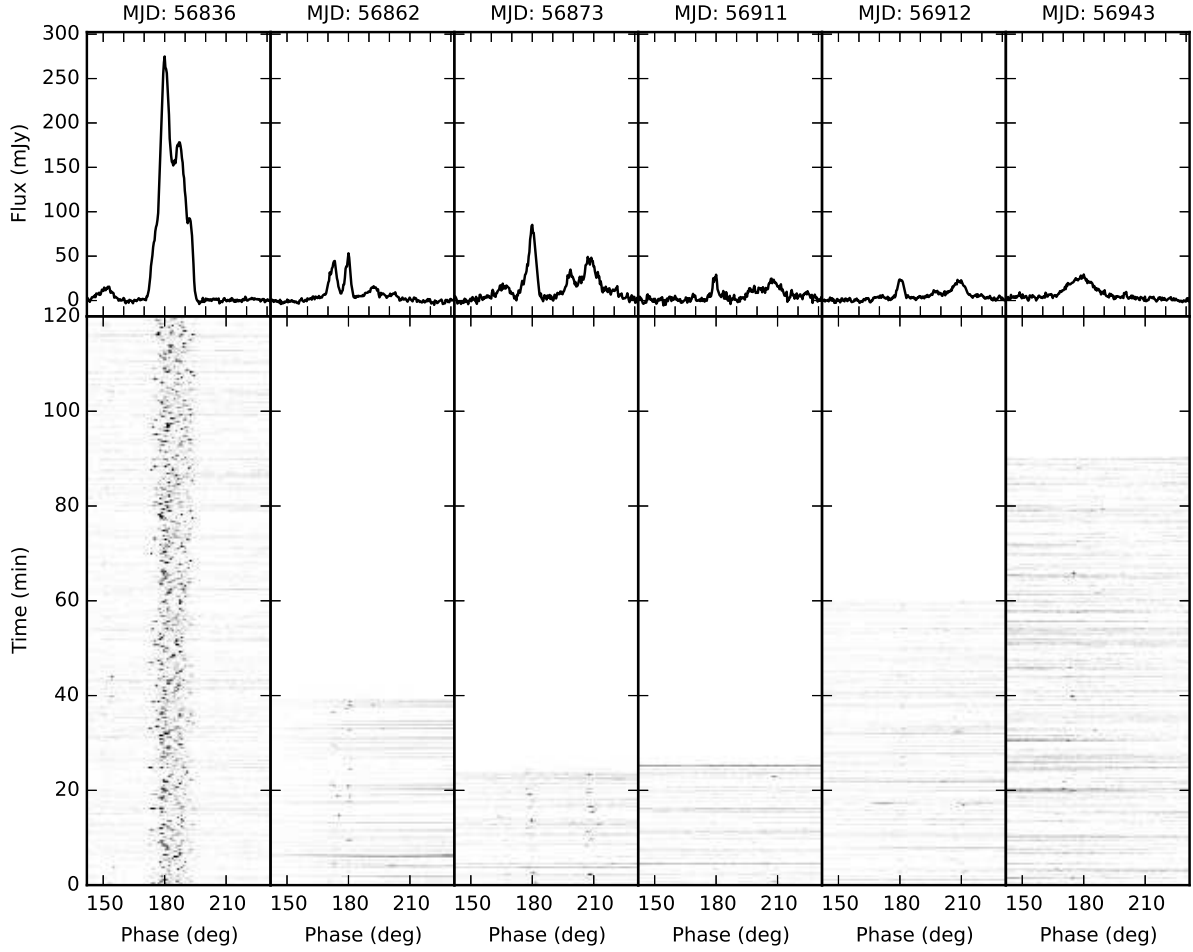}
\caption{Integrated profiles and the phase-time plots of the magnetar
  PSR~J1745$-$2900 on the six days of TMRT observations. The abscissa
  of each subplot is the rotational phase of the pulsar, with just the
  region around the pulse being shown. The MJD for the day of
  observation is given at the top of each subplot. All the
    plots use the same grayscale which is set according to the
    observations of MJD~56911 where the flux density is relatively
    low. The horizontal striping visible in the phase-time plots
    results from low-level radio-frequency interference (RFI).} 
\label{fig:intprofile}
\end{figure*}

The integrated profiles and phase-time plots of the region around the
pulse for each PSR~J1745$-$2900 TMRT observation are shown in
Fig.~\ref{fig:intprofile}. Clearly the pulse profile varies greatly
over the span of the observations, although on the two adjacent days
(MJDs 56911 and 56912) the profiles are the same within the noise.

For the other epochs of PSR~J1745$-$2900 observation, at least two of
calibrator pulsars listed in Table~\ref{tab:fluxpsr} were observed
with the same observation setup.  We obtain flux densities for
PSR~J1745$-$2900 on MJDs 56862, 56873, 56911, 56912 and 56943 of
$2.69\pm0.81$, $3.02\pm1.39$, $0.63\pm0.20$, $0.83\pm0.26$ and
$1.83\pm0.35$~mJy, respectively.  For comparison, the flux density of
this magnetar at the time of its discovery is only 0.2-0.8~mJy
\citep{efk13}. This variation cannot result from diffractive
interstellar scintillation as the characteristic bandwidth is only of
order kHz \citep{sle+14} and so the observations average over many
scintles. The timescale for refractive scintillation is of order years
and hence it is possible that this could contribute to long-term
intensity variations. However, the associated pulse profile changes
(Fig.~\ref{fig:intprofile}) must be intrinsic and it is probable that
the intensity variations are also largely intrinsic. \citet{lak15}
found that the radio emission became stronger and more erratic from
around MJD 56726 and our observations show that this erratic phase continued
till at least MJD 56943. An enhancement of the X-ray emission at the
GC was detected by Swift on MJD~56910 \citep{drm14}, but we found no
obvious change in our radio observations of PSR~J1745$-$2900 on two
subsequent days (MJD~56911, 56912). This X-ray flare is probably
associated with Sgr~A*.

The integrated profile on MJD~56836 shown in Fig.~\ref{fig:intprofile}
is comparable with that seen in GBT observations on MJDs 56751, 56794
and 56808, with a three-peak structure and precursor component. The
precursor component was also detected in the integrated profile of
this magnetar when it was discovered during its X-ray burst phase
(around MJD~56410). But, at that time the main pulse was a single
peak, not three peaks \citep{efk13}. The third component frequently
showing in the GBT observations since MJD~56726 was also detected in
our observations on MJDs 56836, 56862, 56873, 56911 and 56912. All of
these observations suggest that the magnetar was in its erratic state
during our observations. Because we only have six epochs of
observation, it is impossible for us to update the pulsar
timing parameters of this magnetar. In this paper, we focus on
the single-pulse properties of this magnetar.

The phase-time plots in Fig.~\ref{fig:intprofile} clearly show
single-pulse emission spikes located within the phase window
of the integrated profiles. These are more obvious in the plot
corresponding to the MJD~56836 observation because of high
signal-to-noise ratio (S/N) of this observation.  Such spiky
emission is not common in phase-time plots of other pulsars. In the
following section we further discuss the single-pulse properties of
PSR~J1745$-$2900.

\section{Single pulses of PSR~J1745$-$2900}\label{sec:single}
Normally, the system equivalent flux density (SEFD) of TMRT is
about 50~Jy at 8.6~GHz. In our observation, the integration time of
each phase bin is 3.65~ms; therefore only pulses
whose peak exceeds about 30~mJy will be detected at the
1~$\sigma$ level. Since the duty cycle of J1745$-$2900 is less
than 10\%, it is possible to detect single
pulses in the MJD~56836 PSR~J1745$-$2900 observation when its mean
flux density was about 8.75~mJy. But, for other five epochs of observation,
it is hard to detect most of individual pulses except some
strong spiky pulses because of the limited S/N.

\begin{figure}[ht]
\centering
\includegraphics[width=0.6\textwidth,angle=-90]{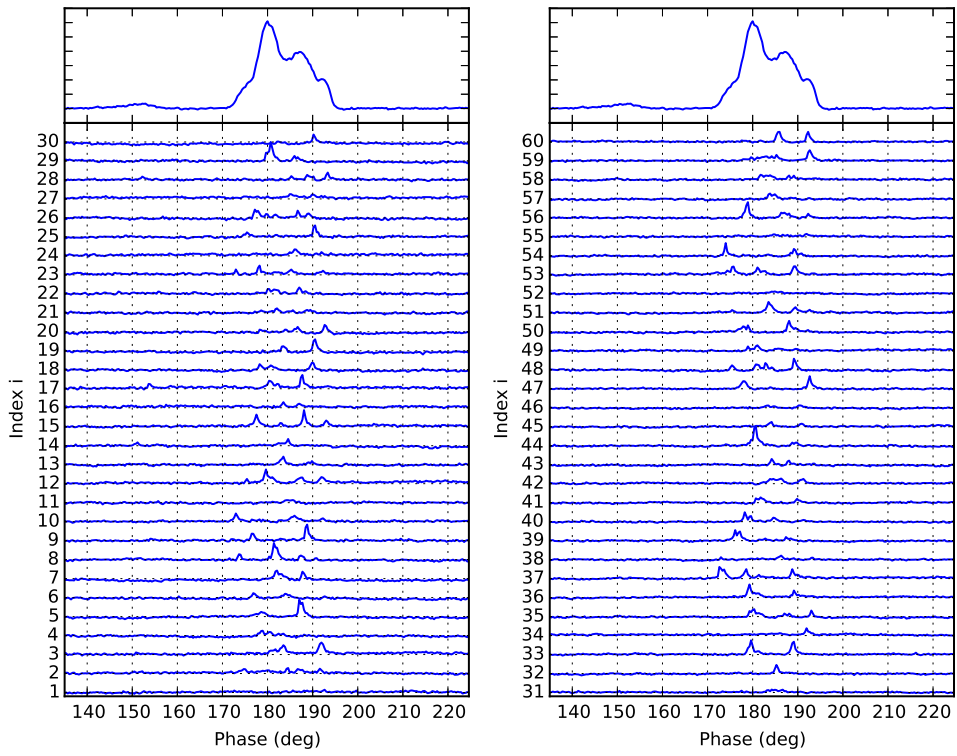}
\caption{Single pulses from PSR~J1745$-$2900 observed on MJD~56836.
The actual pulse number counted from the start of the observation
is $360+\rm i$.}
\label{fig:singlepulse}
\end{figure}

Fig.~\ref{fig:singlepulse} shows two contiguous series of individual
pulses from the observation of MJD~56836. From one period to
the next, the positions of the narrow spikes appear to change
randomly. However, within a given pulse the spikes appear to be
quasi-periodic. PSR~J1745$-$2900 observation with VLA on MJD~56486
indicated that these spikes have a typical separation of
$\sim 10$~ms \citep[cf.,][]{bdd14}. This is discussed further below
in \S\ref{sec:drift}. Even though the spiky pulses are very narrow
compared to the integrated pulse profile, there is evidence in
Fig.~\ref{fig:singlepulse} that some spikes have structure with two or
more peaks or components.

\subsection{The peak flux density distribution of single pulses}
\begin{figure}[ht]
\centering
\includegraphics[width=0.6\textwidth,angle=-90]{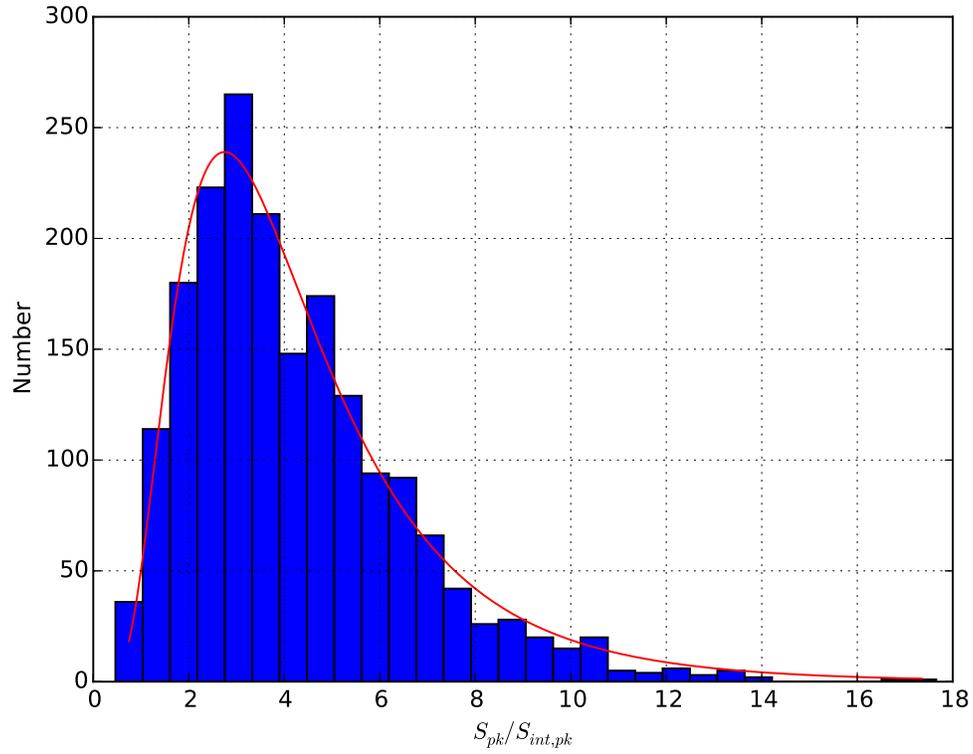}
\caption{Distribution of the peak flux density of single pulses
  relative to the peak flux density of the mean pulse profile for the
  observation of MJD~56836. The best fitting log-normal function is shown as a solid
  curve.}
\label{fig:hiseng}
\end{figure}

Here we present a statistical analysis of the energy distribution
of single pulses based on the observation of MJD~56836.  Following
\citet{lak15}, we normalize the peak flux density in a given pulse,
$S_{\rm pk}$, with the peak flux density of the integrated profile for the
whole observation, $S_{\rm int,pk}$. The histogram of the distribution
of peak flux densities of all the 1913 pulses is shown in
Fig.~\ref{fig:hiseng}. It appears that
the observed distribution follows a log-normal distribution:
\begin{equation}
P_{\rm ln}=\frac{S_{\rm int,pk}}{\sqrt{2\pi} \sigma S_{\rm pk}}
\exp[{-(\ln\frac{S_{\rm pk}}{S_{\rm int,pk}}-\mu)^2 / (2\sigma^2)}]
\end{equation}
where $S_{\rm pk}$ and $S_{\rm int,pk}$ are the peak flux density of a
single pulse and the integrated pulse profile, respectively, and $\mu$
and $\sigma$ are the logarithmic mean and the standard deviation of
the distribution. A Kolmogorov-Smirnov hypothesis test
  \citep{lil67} is used to check whether or not the observed
  distribution can be described as log-normal. The $p$-value of
  log-normality hypothesis test is about 0.07, which is greater than
  the threshold value 0.05, indicating that the observed distribution
  is consistent with a log-normal distribution. The observation data
  are fitted to the log-normal model using a least-squares fitting
  method. The solid curve in Fig.~\ref{fig:hiseng} is the best-fitting
  log-normal distribution with $\mu = 1.34\pm 0.02$ and $\sigma =
  0.57\pm 0.02$.

\begin{figure}[b]
\centering
\includegraphics[width=0.6\textwidth,angle=-90]{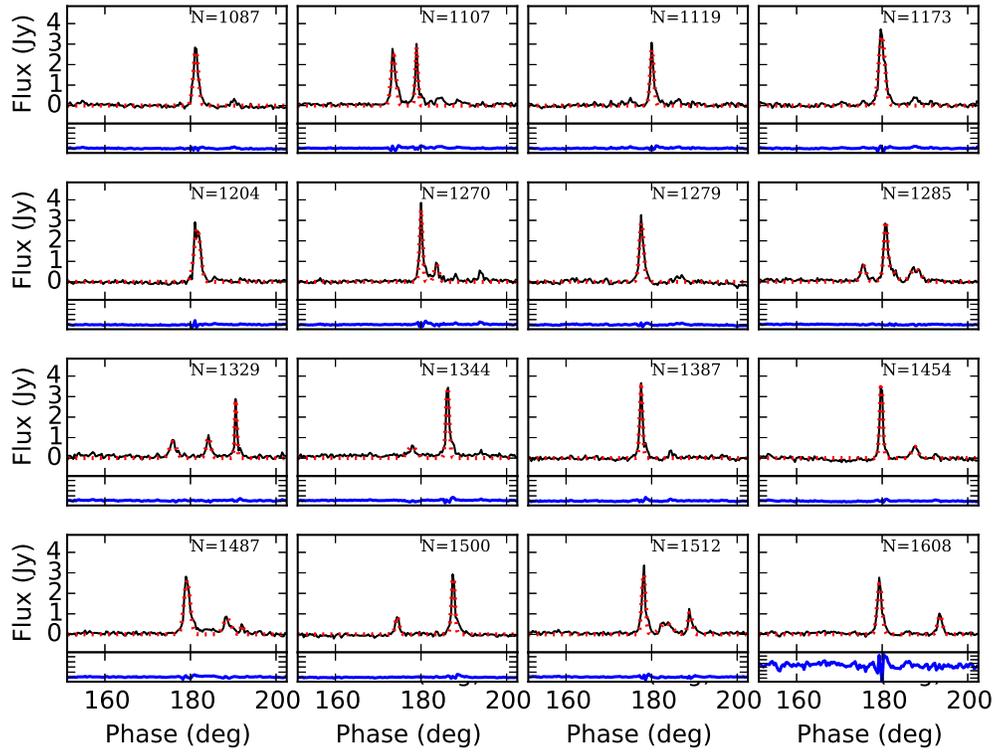}
\caption{Profiles of pulses whose peak flux density exceeds ten
  times the peak flux density of the integrated profile. The observed
  pulsar profiles and fits of Gaussian profiles to the strong features
  are plotted as solid lines and the dotted lines,
  respectively. Pulse numbers are counted from the start of the
  observation. The residuals from the Gaussian fitting are shown
  at the bottom of each sub-panel.}
\label{fig:brightprof}
\end{figure}

In our observation of MJD~56836 we detected 53 bright pulses with peak flux density $S_{\rm pk}$
at least ten times larger than the peak flux density of the integrated profile $S_{\rm int,pk}$.
In order to do further studies on these bright pulse and distinguish
whether or not there are some different properties between these strong pulses compared to
others of lower peak flux density, we also try to characterize
the shape of the 1913 pulses with the Gaussian fitting method. We find that most of
the peak components of these pulses can be fitted with one Gaussian component. The
profile of a sample of the pulses whose peak flux density exceeds ten times that of the
integrated profile is shown in Fig.~\ref{fig:brightprof}, along with the corresponding
Gaussian fitting results.

\begin{figure}[b]
\centering
\includegraphics[width=0.6\textwidth,angle=-90]{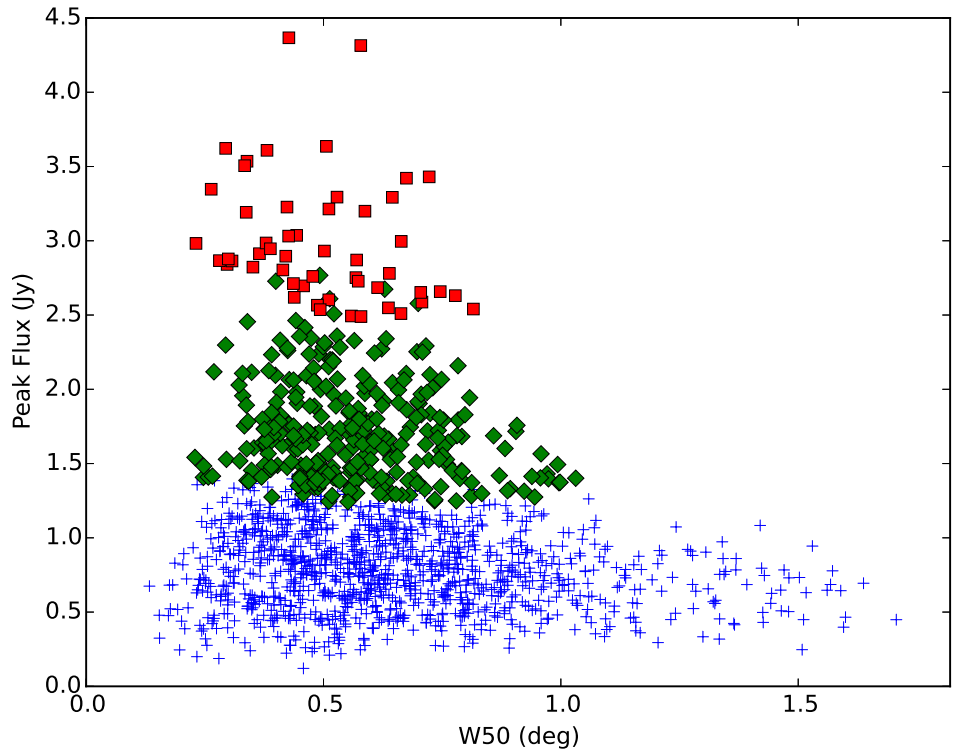}
\caption{Peak flux density versus half-power width for the spikey
pulses. Pulses whose $S_{\rm pk}$ is greater than 10 times $S_{\rm int,pk}$,
in the range of 5 to 10 and less than 5, are labelled with squares, diamonds
and plus markers, respectively.}
\label{fig:pkwidcor}
\end{figure}

With the Gaussian fitting method, we get the width at 50\% of
peak (W50) and fitted peak flux density of single pulses of
PSR~J1745$-$2900. Historical observations of giant pulses indicate
that there is a correlation between the peak flux density and the
width of giant pulses, with the stronger pulses having narrower
pulse widths \citep{jak+98,mnl11}. This phenomenon is predicted by
the giant-pulse radiation model of intensity amplification by induced
Compton scattering in the pulsar magnetosphere plasma \citep{pet04}.
In order to investigate whether this relation exists in the strong
pulses of PSR~J1745$-$2900, we give a scatter plot of the fitted
peak flux density $S_{\rm pk}$ and the W50 width in Fig.~\ref{fig:pkwidcor}.
From the plot, it is clear that the W50 of pulses whose peak flux density
exceeds ten times that of integrated file are in the range of
0.2$^\circ$ to 0.9$^\circ$. For comparison, W50 for the integrated
profile is about 12$^\circ$. The W50 of pulses with lower peak flux
density are in a comparable but somewhat wider range
(0.1$^\circ$ to 1.9$^\circ$). However, there is no clear correlation
between peak flux density and pulse width. Even though the peak flux
density of these pulses is high, their pulse energy is not that large,
as their pulse widths are much narrower than that of the integrated profile.

\begin{figure}[b]
\centering
\includegraphics[width=0.6\textwidth,angle=-90]{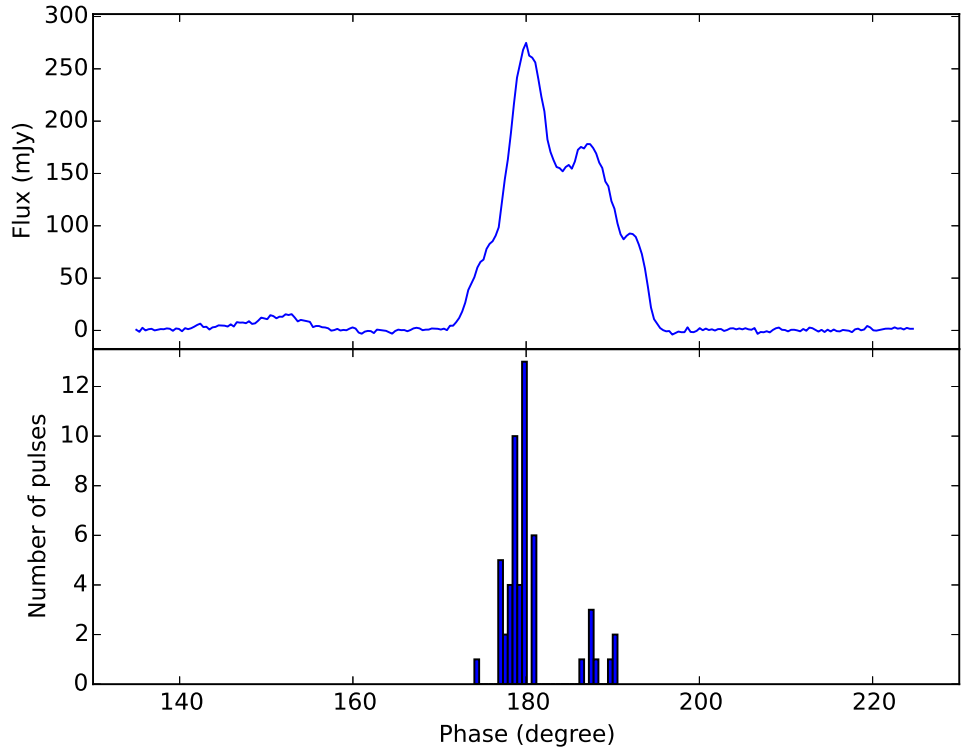}
\caption{Histogram of central pulse phase for the 53 pulses with peak
  flux density greater than ten times the peak flux density of the
  integrated pulse profile. For comparison, the integrated pulse
  profile is shown in the top panel.}
\label{fig:brightphase}
\end{figure}

The distribution in pulse phase of the 53 strong pulses is presented
in Fig.~\ref{fig:brightphase}. This shows that the strong pulses are
preferentially emitted at the phases of the peaks in the integrated
profile. No strong pulses were detected at the phase of the weaker
leading component.

Spiky pulses with peak flux density exceeding ten times that of the integrated
profile are also found in the other five epochs of observation data on PSR~J1745$-$2900.
The number of bright pulses detected in the observation of MJD~56862, 56873,
56911, 56912 and 56943, is 9, 7, 5, 16 and 13 respectively. Most of these
pulses can also be fitted with one Gaussian component with a typical width in
a range of 0.2$^\circ$ to 1.2$^\circ$. They also occur at the pulse phases
of the peaks of the integrated profile.

\subsection{Subpulse drifting}\label{sec:drift}
Many pulsars exhibit drifting subpulses, that is, a progressive shift
of subpulse phase in successive pulses. Subpulse drifting can be
characterized by two numbers: the horizontal separation between
adjacent subpulses in pulse longitude ($P2$) and the vertical
separation between drift bands in pulse periods ($P3$).  Subpulse
drifting can be detected using an auto-correlation analysis
\citep{tmh75} or by Fourier methods \citep{eds03}. Other useful
diagnostics of subpulse modulation are the longitude-resolved standard
deviation (LRSD) and the longitude-resolved modulation index
(LRMI). Useful diagnostic plots from the Fourier methods are the
two-dimensional fluctuation spectrum (2DFS) and the longitude-resolved
fluctuation spectrum (LRFS) which can be used to characterize $P2$ and
$P3$, respectively \citep{wes06, wse07}.

\begin{figure}[b]
\centering
\includegraphics[width=0.9\textwidth,angle=-90]{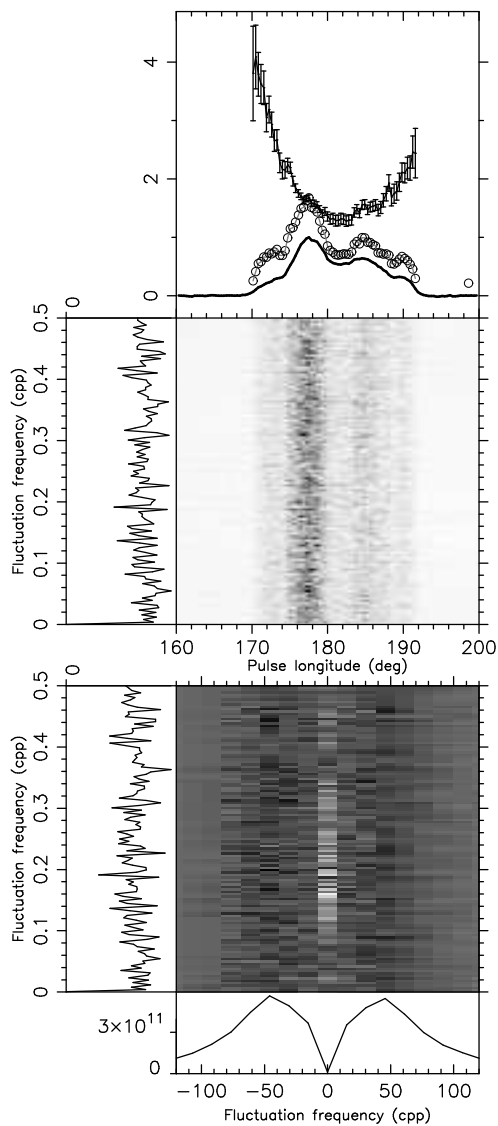}
\caption{Results of subpulse analysis for the PSR~J1745$-$2900
  observation of  MJD~56836. Upper panel: integrated pulse profile (solid line),
  longitude-resolved modulation index (error bars) and standard
  deviation (open circles). Middle panel: longitude-resolved
  fluctuation spectra with the phase-integrated spectrum on the left.
  Lower panel: two-dimensional fluctuation spectrum with integrated
  spectra on the left and below.}
\label{fig:lrfs}
\end{figure}

To investigate the subpulse modulation properties of PSR~J1745$-$2900,
we calculate the LRSD, LRMI, LRFS and 2DFS for the observation of MJD
56835. In order to detect weak features and obtain information
about the trends in an average sense, the input pulse series are divided
into smaller blocks consisting of 256 pulses and so a fast
Fourier transform length of 256 is used in our calculations.
Each of these blocks was analyzed separately and the averaged results are
shown in Fig.~\ref{fig:lrfs}. The upper plot shows a strong pulse modulation
with peaks in the modulation index at the phase of the peaks in the
integrated profile. The middle panel shows that there are no strong
periodicties in the subpulse modulation at any phase or in the
phase-integrated spectrum. Likewise, the 2DFS in the lower plot shows
no significant modulations in the vertical (pulse-to-pulse) direction,
but there are significant broad peaks in the horizontal (pulse phase)
direction. These correspond to the separation of the pulse components
and the preferred phases of the strong pulses of about $8^\circ$ shown
in Fig.~\ref{fig:brightphase}, but do not represent a $P2$ modulation
related to drifting subpulses.

\section{Discussion and conclusions}\label{sec:concl}
Both the flux density and the integrated profile shape of
PSR~J1745$-$2900 show dramatic changes from epoch to epoch in our
observations. This shows that the erratic phase starting around
MJD~56726 seen in the GBT observations \citep{lak15} continued at
least to our last observation on MJD 56943 (2014 Oct. 13). Dramatic
changes in the flux density and integrated profile shape are also
found for other magnetars whose radio radiation has been
detected. Flux density variations observed in magnetar PSR J1550-5418
are up to 50\% on timescales of a few days \citep{crj08}. The flux
density of magnetar PSR J1622$-$4950 varied by up to a factor of
$\sim$10 within a few days and, on average, decreased by a factor of 2
over 700 days \citep{lbb12}. Clearly, the radio radiation of magnetars
is not as stable as other radio pulsars. In normal pulsars the radio
radiation is powered by the rotational energy loss, whereas in
magnetars it is probably powered by the release of the magnetic-field
energy. This difference in energy supply may account for the different
nature of the emission.

To resolve the subpulse properties of PSR~J1745$-$2900, observations
must be made at relatively high radio frequencies to overcome the
effects of interstellar scattering \citep{sle+14}. The TMRT
observations reported in this paper were made in the band 8.2 -- 9.0
GHz with a nominal center frequency of 8.6 GHz. At these frequencies,
the scattering time is of order a few ms. These observations, as well
as previous observations at similar frequencies \citep{lak15,bdd14},
show that the subpulse structure is dominated by strong narrow spikes
with a width of this order. The width of strong pulses in the TMRT
observations ranged between $0.2^\circ$ and $0.9^\circ$, corresponding
to approximately 2 -- 9 ms. It is possible that the observed pulse
widths at frequencies around 8.6~GHz are dominated by interstellar
scattering and that the intrinsic widths of these strong pulses are
narrower. In Fig.~\ref{fig:brightprof}, there is some evidence for
scattering tails on some pulses.  The residuals between the observed
pulses and Gaussian fitting results are often larger at the pulse
tail; see, for example, pulses N=1279, 1329, 1387. Higher-frequency
observations would be required to reveal true intrinsic pulse widths.

Analysis of the peak flux density distribution of the strong pulses
relative to the peak of the integrated profile showed that it was well
described by a log-normal distribution with mean in the logarithm of
1.34 and a standard deviation of 0.57. There is no evidence for a high
energy tail to the distribution. The GBT observations of \citet{lak15}
also showed a log-normal distribution with a similar standard
deviation, but also showed a significant high-energy tail. This
suggests that the pulse intensity distribution may be variable in
time.

From the 1913 pulses observed on MJD~56836 we detected 53 pulses whose
peak flux density exceeds ten times that of the integrated
profile. No correlation was found between the width of the strong
pulses and their peak flux density. Relatively bright spiky pulses
were also detected in the other five epochs of our observation data
with similar properties. The strong narrow pulses
observed from PSR~J1745$-$2900 are not sufficiently strong to be
classed as ``giant'' pulses similar to those detected in some pulsars
such as the Crab pulsar \citep{lcu+95} and PSR~B1937+21
\citep{kt00}. Furthermore, in our observations at least, they do not
have the power-law distribution in amplitude that characterizes giant
pulses and they are not confined to a narrow phase range within the
integrated profile.

For most normal pulsars, the single-pulse radio flux density
distributions are also close to log-normal and most do not emit giant
pulses \citep{cai04,bjb12}. Only three magnetars currently have radio
single-pulse observations and these give different
results. Observations of PSR~J1622-4950 at 3.1~GHz found that the flux
density distribution followed a log-normal distribution
\citep{lbb12}. In contrast, simultaneous single-pulse observation of
the radio emitting magnetar PSR~J1809-1943 at frequencies of 1.4, 4.8
and 8.35~GHz by \citet{ssw09} found that its pulse-energy
distributions were very peculiar, changing on a time scale of days and
not able to be fitted by a single statistical model.

Bright spiky radio emission in the window of the integrated profile
has been detected in a few normal pulsars, for example, PSRs B0656+14
\citep{wsrw06}, B0943+10 \citep{bmr10} and PSR~B0031$-$07
\citep{kss11}, and for two other magnetars, PSR~J1809$-$1943
\citep{ssw09} and PSR~J1622$-$4950 \citep{lbb12}. Strong magnetic
fields at the light cylinder $B_{\rm L}$ and high spin-down luminosity
$\dot{E}$ have been proposed as indicators of giant pulse emission
\citep{cst96, jor02, kni06}. In Table~\ref{tab:spikypulsar} we give
$B_{\rm L}$ and $\dot{E}$ along with pulse period, characteristic age
and surface dipole magnetic field for these six pulsars. Except for
PSR B0031$-$07, these pulsars are comparatively young. The magnetic
field at the light cylinder $B_{\rm L}$ and the spin-down luminosity
$\dot{E}$ are not as high as for pulsars emitting giant pulses and are
in the same range as other normal pulsars. This suggests that the
strong spiky emission seen in magnetars and some normal pulsars is
more closely related to normal radio emission and that the giant
pulses have a different radiation mechanism.

\begin{table}[b]
\centering
\caption{Parameters of pulsars with spiky radio emission}
\label{tab:spikypulsar}
\begin{tabular}{lccccc}
\hline
PSR & $P_{\rm 0}$ & $\tau_{\rm C}$ & $\dot{E}$ & $B_{\rm S}$ & $B_{\rm L}$ \\
  & (s) & (yr) & (ergs/s) & (G) & (G) \\
\hline
B0031$-$07   &  0.94  &   3.66$\times 10^7$ & 1.92$\times 10^{31}$ &  6.27$\times 10^{11}$ &  7.02$\times 10^0$ \\
B0656+14     &  0.38  &   1.11$\times 10^5$ & 3.81$\times 10^{34}$ &  4.66$\times 10^{12}$ &  7.66$\times 10^2$ \\
B0943+10     &  1.10  &   4.98$\times 10^6$ & 1.04$\times 10^{32}$ &  1.98$\times 10^{12}$ &  1.40$\times 10^1$ \\
J1622$-$4950 &  4.33  &   4.03$\times 10^3$ & 8.29$\times 10^{33}$ &  2.74$\times 10^{14}$ &  3.18$\times 10^1$ \\
J1745$-$2900 &  3.76  &   4.31$\times 10^4$ & 1.02$\times 10^{33}$ &  7.30$\times 10^{13}$ &  1.28$\times 10^1$ \\
J1809$-$1943 &  5.54  &   1.13$\times 10^4$ & 1.80$\times 10^{33}$ &  2.10$\times 10^{14}$ &  1.16$\times 10^1$ \\
\hline
\end{tabular}
\end{table}

\section*{Acknowledgments}
This work was supported in part by the National Natural Science Foundation of China (grants 11173046 and 11403073),
Natural Science Foundation of Shanghai No.~13ZR1464500, National Basic Research Program of China (973 program)
No.~2012CB821806, the Strategic Priority Research Program ``The Emergence of Cosmological Structures" of the
Chinese Academy of Sciences, Grant No.~XDB09000000 and the Knowledge Innovation Program of the Chinese Academy
of Sciences (Grant No.~KJCX1-YW-18) and the Scientific Program of Shanghai Municipality (08DZ1160100).


\label{lastpage}


\end{document}